# Boosting spintronics with superconductivity

Guang Yang[1,*], Chiara Ciccarelli[2,**], Jason W. A. Robinson[1,†]

1. Department of Materials Science & Metallurgy, University of Cambridge, Cambridge, UK
2. Cavendish Laboratory, University of Cambridge, Cambridge, UK

[*] gy251@cam.ac.uk
[**] cc538@cam.ac.uk
[†] jjr33@cam.ac.uk

**Abstract**

Spintronics aims to utilize the spin degree of freedom for energy-efficient, non-volatile memory and logic devices. In this research update, we review state-of-the-art developments and new directions in charge- and spin-based memory/logic with a focus on spintronics and the fascinating potential for superconductivity to boost spin transmission via spin-polarized quasiparticles or triplet Cooper pairs.

## Introduction

Despite efforts to reduce the energy required for electrically-driven magnetization switching (data writing) of ferromagnetic elements, a large energy gap between data writing and retention remains a major challenge for energy-efficient spintronics. Fundamentally, the high energy dissipation of magnetization switching is related to Ohmic losses associated with spin transmission via charge currents. Identifying mediums to carry spin currents with low Ohmic losses is therefore key to the development of spintronic memory.

The emerging field of superconducting spintronics offers new concepts for spin transmission by combining superconducting phase coherence and magnetism.[1] Ferromagnetism and conventional (s-wave, spin singlet) superconductivity do not coexist in bulk materials due to their opposing properties. However, spin-related phenomena can emerge at s-wave superconductor/ferromagnet (SC/FM) interfaces:[1-4] for example, spin can transmit within a superconductor to micrometre length scales, supported by spin-polarized quasiparticles that occupy the energy states above the superconducting energy gap.[5,6] The transmission length in this case, much longer than the superconducting (singlet) coherence length $\xi_s$, is broadly determined by the spin-orbit and spin-flip scattering lengths. Alternatively, spin can be carried by the superconducting state itself via spin-polarized triplet Cooper pairs where the transmission distance is determined by $\xi_s$.[2] Compared with spin-polarized quasiparticle currents, equilibrium spin-polarized triplet supercurrents are dissipationless and phase coherent.

## 1 Development of energy-efficient spin-information writing

Initiated by the 4 Mb memory in 2006, magnetic random access memory (MRAM) has rapidly expanded its memory capacity to 1 Gb by 2019.[7] MRAM has overcome numerous hurdles including the information writing process. The first generation of (field-driven) MRAM operates using orthogonal magnetic fields generated by current lines to switch the magnetization. In this case scalability is limited to a 90-nm-dimension technology node.[8] The second generation of (current-driven) MRAM uses spin-polarized charge currents to switch the magnetization of a FM element through *s-d* orbital exchange interaction between the conduction electron spin and the localized magnetic moment, i.e., spin-transfer torque (STT).[9] The development of STT-MRAM enabled further scalability via a localized writing approach, hence the switching energy averaged in a single bit is reduced. However, STT-MRAM struggles to reduce the writing time due to the associated thermal incubation delay time for STT[10], which increases in energy consumption and limits the operation speed.

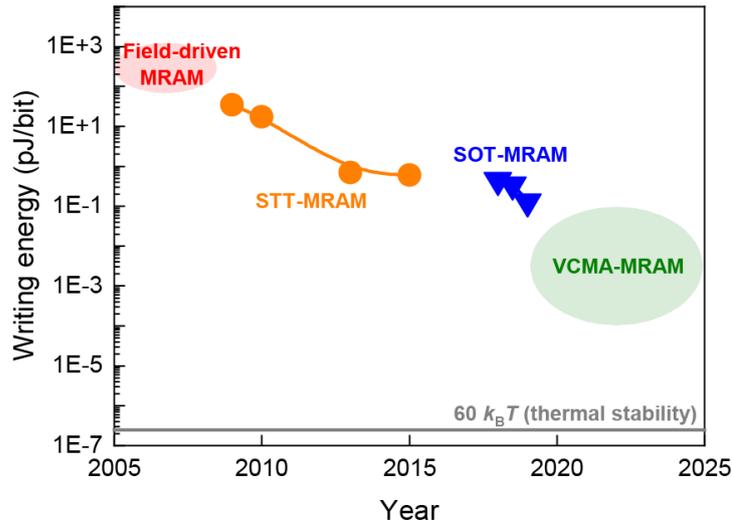

**FIG. 1.** Writing energy reduction trend of spin-based memories. Performance data from CMOS-integrated memory cells at the industry level. STT-MRAM data from refs.[11-14] and SOT-MRAM data from refs.[15,16].

Spin-orbit torque (SOT) switching, requiring materials with high spin-orbit coupling (SOC) as the spin current source,[17] led to a reduction in the switching time to sub-nanosecond timescales[18] with the switching energy reduced by three orders of magnitude in comparison to the original field-driven MRAM (Fig. 1). However, the energy required to maintain the magnetic information is largely determined by the thermal stability energy, about 60 $k_BT$ (where $k_B$ is the Boltzmann constant and $T$ is the temperature), which translates into an energy gap of five orders of magnitude between data writing and retention. This gap mainly arises from Ohmic losses associated with current-writing in both STT and SOT. Different strategies have been proposed in recent years to reduce this gap. For example, in the voltage-controlled magnetic anisotropy (VCMA) approach,[19,20] a voltage is used to change the magnetic anisotropy and make the magnetization precessionally unstable, promoting switching.[20] VCMA writing, even assisted by STT or SOT[21], has a faster energy consumption scalability and promises sub-fJ switching energies below a 30-nm-dimension technology node[22] and hence is seen as the next generation of MRAM.

## 2 Materials with high charge-spin conversion efficiencies and low switching power consumption

While STTs are limited by the spin polarization of the FM injector, SOTs can in principle be increased to much higher values. From a mechanism point of view, SOTs mainly arise from the bulk spin Hall effect (SHE),[23] the interfacial Rashba effect,[24] or spin-momentum locking.[25] In general terms, the SHE originates from 5d heavy metal materials with strong SOC such as Pt,[26] Ta,[17] and W;[27] the interfacial Rashba effect is induced at interfaces with strong SOC and broken inversion symmetry such as Bi/Ag[28] and SrTiO$_3$/LaAlO$_3$[29] interfaces; and the spin-momentum locking arises from the topologically protected surface states in topological insulators (TIs) such as (BiSb)$_2$Te$_3$,[30] BiSe,[31,32] and BiSb.[33] High charge-spin conversion efficiencies have also been reported in non-magnetic oxides such as W(O)[34], antiferromagnets such as (001)-IrMn$_3$,[35] and transition-metal dischalcogenide (TMD) such as WTe$_2$[36]. In Table I, we have summarized the reported values of the charge-spin conversion efficiency $\xi_{SOT}$ at room temperature and the corresponding longitudinal resistivity $\rho_{xx}$ for different spin source materials. We note that, even for the same material, achieving consistent values of $\xi_{SOT}$ is difficult when using different measurement techniques. Nevertheless, $\xi_{SOT}$ is enhanced in alloyed heavy metals and materials with unconventional band structures such as TIs and TMDs.

**TABLE I.** Summary of the longitudinal resistivity $\rho_{xx}$, charge-spin conversion efficiency $\xi_{SOT}$, and measurement techniques for different spin source materials at room temperature.

| Spin sources | $\rho_{xx}$ (μΩ·cm) | $\xi_{SOT}$ | Measurement techniques |
| --- | --- | --- | --- |
| **Heavy Metal** | | | |
| Ta | 190 | 0.12 | DC[17] |
| Pt | 22.7 | 0.07 | ST-FMR[36] |
|  | 27.4 | 0.16 | DC[37] |
| W | 185.7 | 0.14 | DC[38] |
| **Pt-based alloy** | | | |
| Pt$_{57}$Cu$_{43}$ | 82.5 | 0.44 | Harmonic[37] |
| Pt$_{75}$Pd$_{25}$ | 57.5 | 0.26 | Harmonic[39] |
| Pt$_{70}$(MgO)$_{30}$ | 58 | 0.28 | Harmonic[40] |
| $Pt_{75}Au_{25}$ | 83 | 0.3 | Harmonic[41] |
| **Antiferromagnet** | | | |
| $IrMn_3$ (001) | 160 | 0.2 | ST-FMR[35] |
| $IrMn_3$ (111) | 198 | 0.12 | ST-FMR[35] |
| PtMn | 180 | 0.11 | Harmonic[42] |
| **Topological insulator** | | | |
| $(Bi, Sb)_2Te_3$ | 4020 | 0.4 | DC[31] |
| $Bi_2Se_3$ | 1060 | 0.16 | DC[31] |
| | 4117 | 1.75 | ST-FMR[32] |
| BiSb | 400 | 52 | DC[33] |
| $Bi_xSe_{1-x}$ | 7143 | 18.6 | DC[43] |
| **Transition-metal dischalcogenide** | | | |
| $WTe_2$ | 580 | 0.51 | ST-FMR[36] |
| $MoTe_2$ | 550 | 0.03 | ST-FMR[44] |
| $PtTe_2$ | 95 | 0.15 | ST-FMR[45] |

We also compare the power consumption defined as $P = \rho_{xx} (J_c)^2$, where $J_c$ is the critical current for the spin source material to switch an adjacent FM layer. We select the materials listed in TABLE I for which the switching current $J_c$ is quoted in the literature and plot the switching power consumption *vs* longitudinal resistivity in Fig. 2. As shown in Fig. 2, $WTe_2$ has the lowest switching power consumption, down to $5\times10^{10}$ mW·cm$^{-3}$, three orders of magnitude lower than Pt. Pt-Cu alloys with low resistivity show a relatively low switching power consumption down to $5\times10^{11}$ mW·cm$^{-3}$. We note that these values do not take into account current shunting effects.

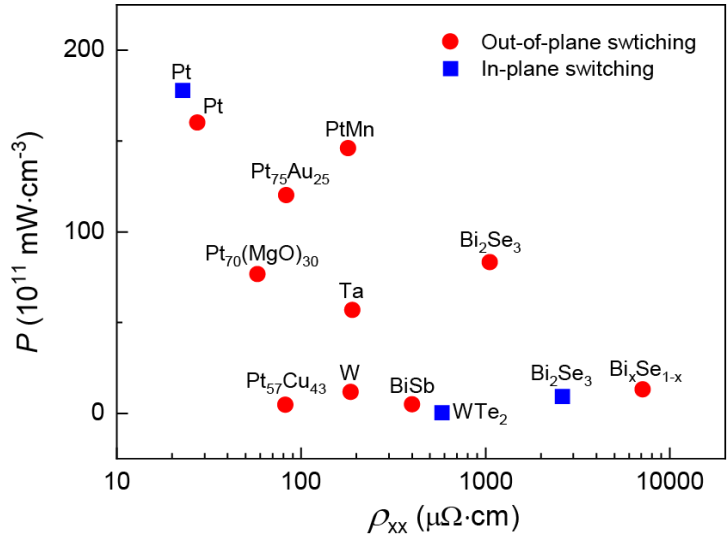

**FIG. 2.** Benchmarking of room-temperature switching power consumption $P$ of SOT devices *vs* longitudinal resistivity $\rho_{xx}$ for different spin source materials. Out-of-plane and in-plane switching mean that the spin source material was applied to switch an adjacent ferromagnetic layer with perpendicular magnetic anisotropy and in-plane magnetic anisotropy, respectively. Data are taken from TABLE I.

# 3 Energy-efficient computing beyond von Neumann architecture

In the von-Neumann architecture, the physical separation between processor and memory and the continuous exchange of data between these two components inevitably limits the data transfer bandwidth (the "memory wall") and increases energy consumption in the interconnecting circuitry (the "power wall"). To overcome these limits, alternative schemes to the von-Neumann architecture have been proposed.

The "in-memory computing" scheme aims to blur the traditional boundaries between logic and storage by allowing the memory unit to perform computational tasks, leading to improvements in the computational efficiency. This is potentially useful in data-centric applications such as machine learning and scientific computing, and reduces energy consumption.[46]

Spintronics offers different "in-memory computing" solutions from probabilistic[47] to neuromorphic computing[48]. These device architectures rely on the electrical manipulation of the magnetization of FM elements and hence, it is crucial to achieve this with the lowest possible energy (broadly determined by the anisotropy energy for magnetization reversal).

In superconductors, charge flow is dissipationless. However, ferromagnetism and s-wave (singlet) superconductivity are, in general, incompatible since singlet Cooper pairs have antiparallel spins and hence, the superconducting state does not support spin accumulation. Recent scientific advances in this field suggest that this is not necessarily the case if singlet pairs can convert to a spin-polarized triplet state at carefully engineered SC/FM interface, opening fascinating possibilities for low-energy-consumption spintronics by integrating superconducting elements. In what follows we will review these advances.

# 4 From conventional spintronics to superconducting spintronics

The prediction of charge-spin separation in a superconductor[49] initiated studies on spin-related phenomena in superconductors. Although these have mainly focused on quasiparticles,[50,51] a complete synergy between superconductivity and magnetic order is possible through the generation of equal-spin triplet Cooper pairs at s-wave SC/FM interfaces.[1-4] In this section, we discuss both equilibrium and non-equilibrium spin transmission in the s-wave superconductors, and highlight the recent discovery of pure spin supercurrents.

## 4.1 Non-equilibrium spin transmission in superconductors

According to the Bardeen–Cooper–Schrieffer theory of superconductivity, the superconducting gap $\Delta$ is related not only to the strength of the pair correlation, but also to the presence of quasiparticle excitations and their energy distribution. Quasiparticles can be excited thermally and electrically, which corresponds to different modes of the distribution function i.e., the energy and the charge modes, respectively. When the excitation is chargeless (e.g., radiation), only the energy is transferred to the superconductor. If instead a charged particle is injected into the superconductor, the balance between electron- and hole-like excitation is broken, resulting in a charge imbalance.[52] The charge imbalance relaxes into an equilibrium state depending on the lifetime of quasiparticles. The injection of spin-polarized currents or spin currents will result in a spin-charge separation in the superconductor because the charge imbalance and the spin imbalance relax independently (predicted by Kivelson and Rokhsar[49] and later observed[5,6]).

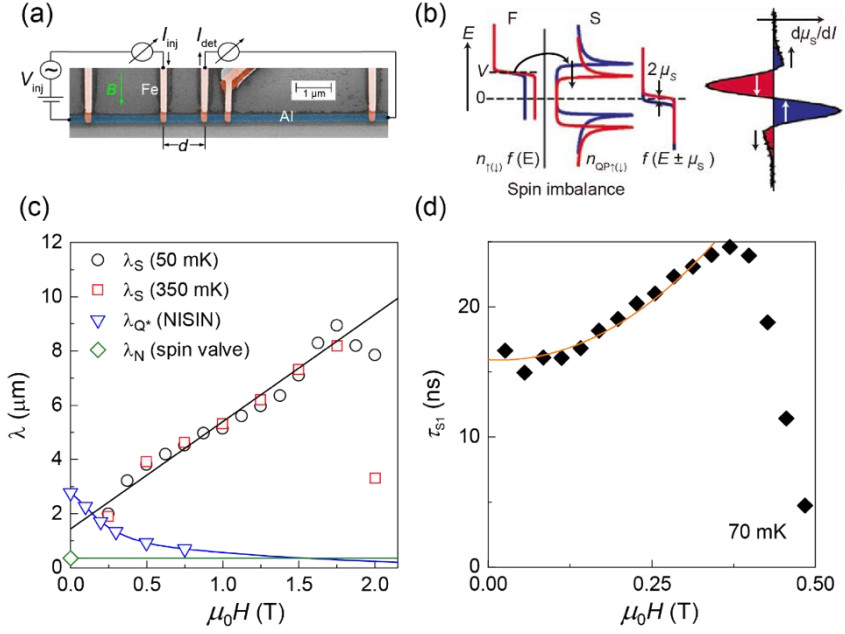

**FIG. 3.** Experimental demonstration of spin-charge separation in superconducting Al by Hübler *et al.*[5] and Quay *et al.*[6] (a) Scanning electron micrograph of the lateral spin-valve device for non-local measurements used by Hübler *et al.*[5]. (b) An in-plane magnetic field ($B$) spin-splits the superconducting density of states in Al, which suppresses the spin-relaxation of quasiparticles.[6] (c) The spin-relaxation length $\lambda_S$ and the charge imbalance length $\lambda_Q$ vs $B$.[5] In the large field regime, $\lambda_S \geq \lambda_Q$, spin-charge separation is obtained. (d) Spin relaxation time $\tau_{s1}$ vs $B$.[6]

Figure 3 shows experimental results confirming spin-charge separation in a lateral spin-valve device with superconducting Al and spin currents injected from a FM layer of Fe. When an external field $B$ is applied (in-plane), the superconducting density of states (DOS) of Al is spin-split [Fig. 3(b)] causing a suppression of spin-flip scattering, resulting in an increase in the spin-relaxation length $\lambda_S$. Compared with the charge imbalance length $\lambda_Q$, $\lambda_S \geq \lambda_Q$ is a signature of the spin-charge separation.

It is important to note that the above results are difficult to interpret without considering the exchange field induced spin-splitting in the superconductor. According to the initial theory and experiments on spin injection into superconductors, the spin-relaxation length in the superconducting state was assumed to be comparable to or lower than that in the normal state, depending on the different spin relaxation mechanisms.[53,54] Recently, a non-equilibrium theory of superconductors with a spin-splitting field has been established, where a general description of non-equilibrium modes including charge, energy, spin, and spin energy is provided.[55,56] The coupling between these modes is the key for understanding long range spin relaxation in the superconducting state [Fig. 3(c)], but also spin-dependent thermoelectric effects including the spin Seebeck effect in superconductors.

Pure spin detection by a superconductor with strong SOC has also been demonstarted.[57-59] In this case, the quasiparticle-mediated inverse SHE converts the injected spin current $J_S$ into a charge current $J_Q$ with the expected relation $\mathbf{J_Q} \sim \mathbf{J_S} \times \mathbf{s}$, where $s$ is the spin polarization unit vector. In lateral spin-valve structures, the strong attenuation of the inverse spin Hall signal when the distance between the voltage probes exceeds the quasiparticles' charge imbalance length confirmed the role of quasiparticles in the spin-charge conversion process. The strong enhancement reported recently of the inverse SHE below the superconducting transition temperature was explained with the increase of the spin conductivity when the quasiparticles' DOS is exchange-split.[59]

## 4.2 Equilibrium spin transmission across proximitized SC/FM interfaces

The first observations of the superconducting proximity effect dates back to 1932 when R. Holm and W. Meissner found it was possible for two superconductors separated by a normal metal (NM) to have zero resistance.[60] The proximity effect at a SC/NM interface can be understood via a double-charge transfer process, i.e., Andreev reflection.[61] The Andreev process retains phase coherence, hence the leakage of superconducting pair correlations into the NM can decay over a relatively long distance, determined by disorder scattering and temperature. If the NM is replaced by a FM, the Andreev process changes. Due to the exchange energy induced momentum mismatch between the majority and minority spin electrons at the Fermi energy in the FM, the proximitized singlet Cooper pairs acquire a finite centre-of-mass momentum.[62,63] The Cooper pair correlations in the FM decay exponentially fast with an oscillatory-dependence on FM layer thickness, resulting in a much shorter transmission distance in the FM compared to a NM.

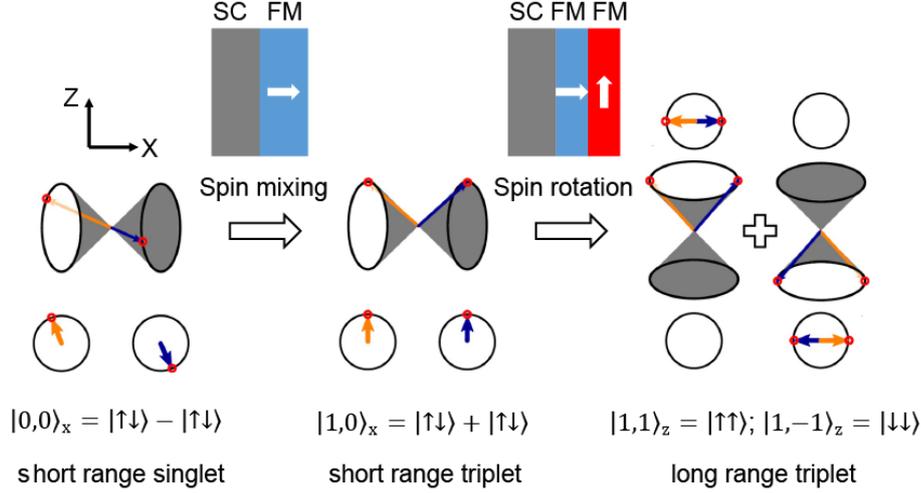

**FIG. 4.** Cooper pair conversion between s-wave spin-singlet and s-wave spin-triplet states. Spin-mixing at the proximitized SC/FM interface shifts the spin-dependent phase of electrons in a singlet state, generating zero-spin triplet pair correlations. A second interface with a misaligned magnetization (here, the orthogonal magnetized configuration is illustrated for simplicity) leads to spin rotation with the quantization axis from $S_x$ to $S_z$, and hence, equal-spin triplet Cooper pairs form and slowly decay in the FM.

In 2001, Bergeret *et al.*[64] predicted that spin-triplet pair correlations are induced in a s-wave spin-singlet superconductor in contact with a magnetically inhomogeneous FM. Fig. 4 illustrates the mechanism of pair conversion between spin-singlet and spin-triplet states, where the ferromagnetic bilayer has misaligned magnetizations.[65] Of the three spin-triplet components, only equal-spin triplets with the z-component $m_s = \pm 1$ can penetrate far into a homogenous FM with its magnetization along the z-axis. This is because both paired electrons are in the same spin band, either the majority band for $m_s = 1$ or the minority band for $m_s = -1$.

Over the past few years, the control of equal-spin triplet pair creation and spin-polarized supercurrents has been widely studied in SC/FM/SC Josephson junctions and SC/FM1/FM2 triplet spin valves. In SC/FM/SC junctions, various spin-mixing layers have been added to the symmetric SC/FM interfaces to demonstrate long-range triplet components (LRTCs), including rare-earth magnetic spirals,[66] antiferromagnets,[67] Heusler alloys,[68] and transition-metal ferromagnets.[69,70] Similarly, in SC/FM1/FM2 triplet spin valves, the variation of the superconducting transition temperature ($T_c$) induced by triplets creation has been successfully demonstrated by controlling the magnetization misalignment between FM1 and FM2 layers in spin valves.[71,72]

Apart from using complex magnetic structures to create equal-spin triplets, recent theories propose interfacial SOC (ISOC) that can also function as a spin-mixing layer for pair conversion.[73] Banerjee *et al.* measured a field-dependent $T_c$ change in Nb/Pt (t ≤ 2 nm)/Co/Pt multilayers in which the Pt is assumed to break structural inversion symmetry inducing a Rashba spin-orbit field.[74] The results show an unconventional suppression of $T_c$ with in-plane and out-of-plane magnetic fields, suggesting that the creation of equal-spin triplet pairs can be controlled by the magnetic orientation of a single magnetically homogeneous FM layer. Another theoretical and subsequent experimental work demonstrated that the superconducting condensation energy can induce the magnetic reorientation in a ISOC-present SC/FM heterostructure.[75,76] We also note that vertical heterostructures with in-plane magnetization and ISOC interaction have been experimentally explored, but LRTC due to ISOC have not been observed.[77,78] Based on previous theoretical works,[73] the conditions for the generation of LRTC by SOC are quite restrictive in vertical structures, more favourable layouts are involving lateral structures where supercurrents have a component flowing in the direction parallel to the hybrid interfaces.[75,79]

### 4.3 Experimental evidence of pure spin supercurrents—A systematic spin pumping characterization study

Although the existence of spin-aligned triplet supercurrents is well established, the spin degree of freedom of such supercurrents is only inferred from supercurrent measurements, i.e., the observation of LRTC necessarily implies spin transmission at the proximitized SC/FM interfaces. The direct observation of a spin-polarized supercurrent is still lacking. In the superconducting state, the electrical characterization of spin-charge conversion by conventional methods is challenging due to the zero resistance, which shunts any voltage-based measurement. It has been proposed that a charge supercurrent can induce a transverse equilibrium spin current in analogy to the conventional SHE.[80] This superspin Hall effect and the corresponding inverse effect, namely the inverse superspin Hall effect, give rise to a phase shift in the Josephson junction, which can be used to detect the spin polarization of a supercurrent carried by triplet Cooper pairs.[81] On the other hand, spin pumping is a mature technique used in non-superconducting spintronics to perform dynamic spin injection.[82] A recent study showed that spin can be pumped to a superconducting NbN film via quasiparticles.[83]

Based on the theory of SOC-induced triplets generation,[73] we have designed a series of sample stacks and conducted systematic experiments to establish the existence of pure spin supercurrents via spin pumping measurements. Two sample stacks were compared as shown in Fig. 5(a). In both structures, the 6-nm-thick Py layer is sandwiched between two Nb layers of equal thickness to enhance the spin pumping efficiency out of Py through two interfaces. In layout 2, additional Pt layers were grown beyond the Nb layers as efficient spin sinks due to their strong SOC. The samples were mounted upside down on a coplanar waveguide, and the transmitted microwave power was measured with a vector network analyser in the presence of an in-plane external magnetic field lower than the superconducting critical field ($H_{c2}$) of Nb. A key parameter in these measurements is the ferromagnetic resonance (FMR) linewidth, directly proportional to the damping, which describes the spin angular momentum loss via spin precession.

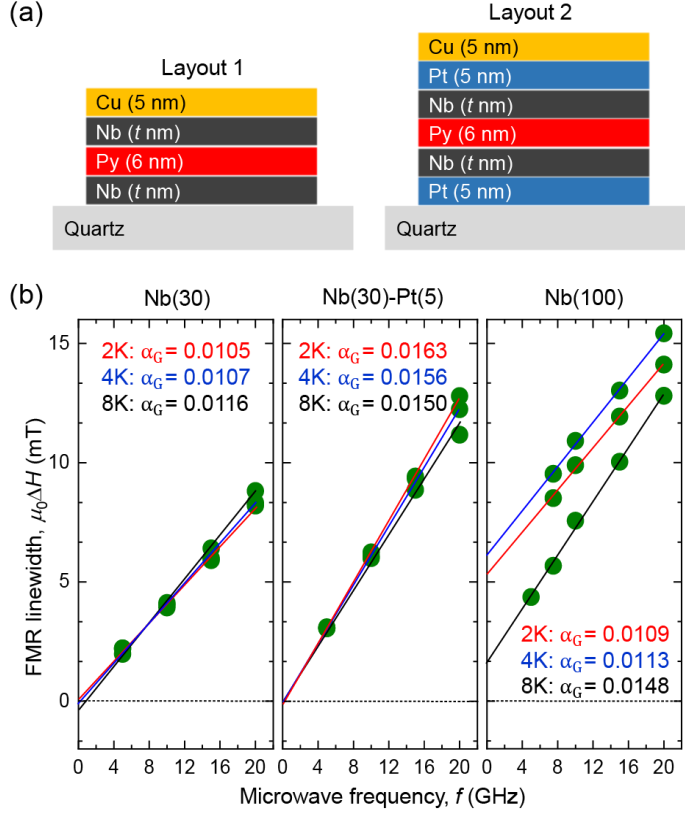

**FIG. 5.** (a) Two layouts of heterostructures with and without SOC layer (Pt) for spin pumping experiments. (b) Frequency dependence of the FMR linewidth above and below $T_c$. The left and right panels correspond to layout 1 with $t_{Nb}$ = 30 nm ($T_c$ = 5.7 K)[84] and $t_{Nb}$ = 100 nm ($T_c$ = 8.9 K),[85] respectively, and the central panel corresponds to layout 2 with $t_{Nb}$ = 30 nm ($T_c$ = 5.8 K).[84]

Fig. 5(b) shows the frequency dependence of the linewidth in three different samples for temperatures above and below $T_c$. In the samples with thin Nb (30 nm),[84] the inhomogeneous (frequency-independent) contribution is small and shows a negligible dependence on temperature, while for thicknesses of Nb comparable to the London penetration depth ($\lambda_L$=100 nm), Meissner screening is seen in a much larger temperature variation that should be taken into account when studying the temperature-induced variations of the FMR linewidth.[85]

In Fig. 6 the FMR linewidth at 20 GHz is plotted for the two layouts as a function of Nb thickness $t_{Nb}$ at different temperatures. In both graphs, $t_{Nb}$ is kept much below $\lambda_L$ hence any effect connected to Meissner screening can be ignored. Above $T_c$, Nb is a NM spin sink and the graphs are all well fitted with spin pumping theory based on spin diffusion.[86] Below $T_c$ the two types of samples behave very differently. In the Pt-absent heterostructures, the data are well described by existing theory. The extracted temperature dependence of the spin-mixing conductance and spin diffusion length reflects two effects: the reduction of spin injection efficiency ascribed to the Py-Nb band structure mismatch below $T_c$ and the shorter quasiparticle lifetime before decaying into Cooper pairs via Andreev reflection. Both observations support a model in which spin transmission involves the states above the superconducting gap and hence is mediated by quasiparticles.[87] In layout 2, the presence of the additional Pt layer causes a deviation from known models, and the data cannot be described (fitted) with existing theory. Interestingly, for $t_{Nb}$ < 40 nm, the FMR linewidth increases below $T_c$, reaching a maximum at $t_{Nb}$ = 30 nm that exceeds the normal state. This suggest a more efficient spin pumping in the superconducting state of Nb, even more efficient than Pt itself ($t_{Nb}$ = 0 nm) in the normal state.

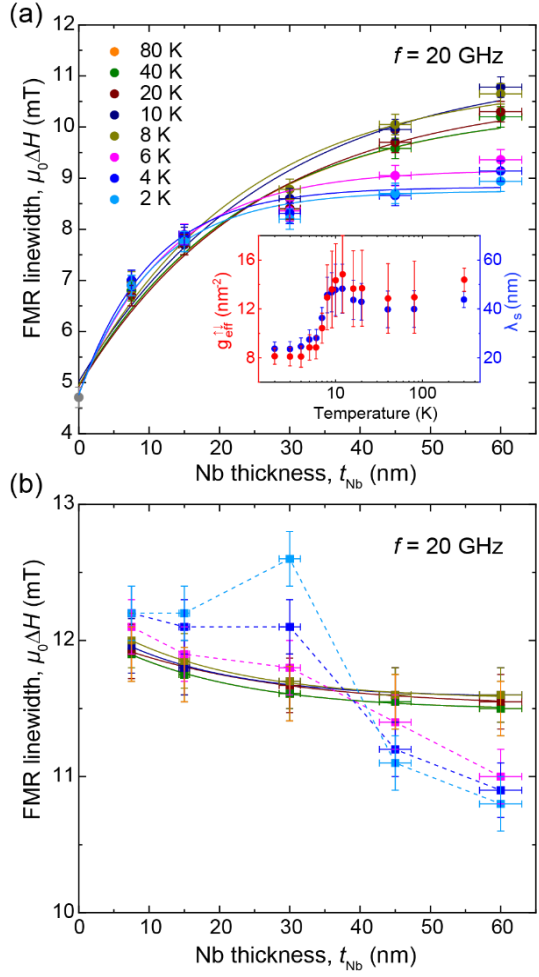

**FIG. 6.** (a) FMR linewidth *vs* Nb thickness for layout 1 [shown in Fig. 5(a)] at various temperatures. The solid lines are fitted to estimate the effective values of spin-mixing conductance ($g_{eff}^{\uparrow\downarrow}$) and the spin diffusion length ($\lambda_S$), as plotted in the insert. (b) FMR linewidth as a function of Nb thickness for layout 2 [shown in Fig. 5(a)] at various temperatures.[84]

To pin-down the role of the Nb-Pt interface, Pt was substituted with different spin sinks with a fixed 5-nm-thickness.[84] The increase in damping below $T_c$ is only observed for Pt, Ta and W but not for antiferromagnetic $Fe_{0.5}Mn_{0.5}$ or Cu. This excludes the quasiparticle-contributed processes such as crossed Andreev reflection and elastic co-tunneling, which play a role in dissipating spin angular momentum away from the Py layer.[88] According to the theory,[73,89] SOC in combination with a magnetic exchange field $h_{ex}$ can generate equal-spin triplet Cooper pairs at a single magnetically homogeneous SC/FM interface. Hence, the enhanced spin pumping at the superconducting state can be explained by the existence of pure spin supercurrents, which flow across the Nb layers to the spin sink layers. This has been further tested by two additional experiments, described below.

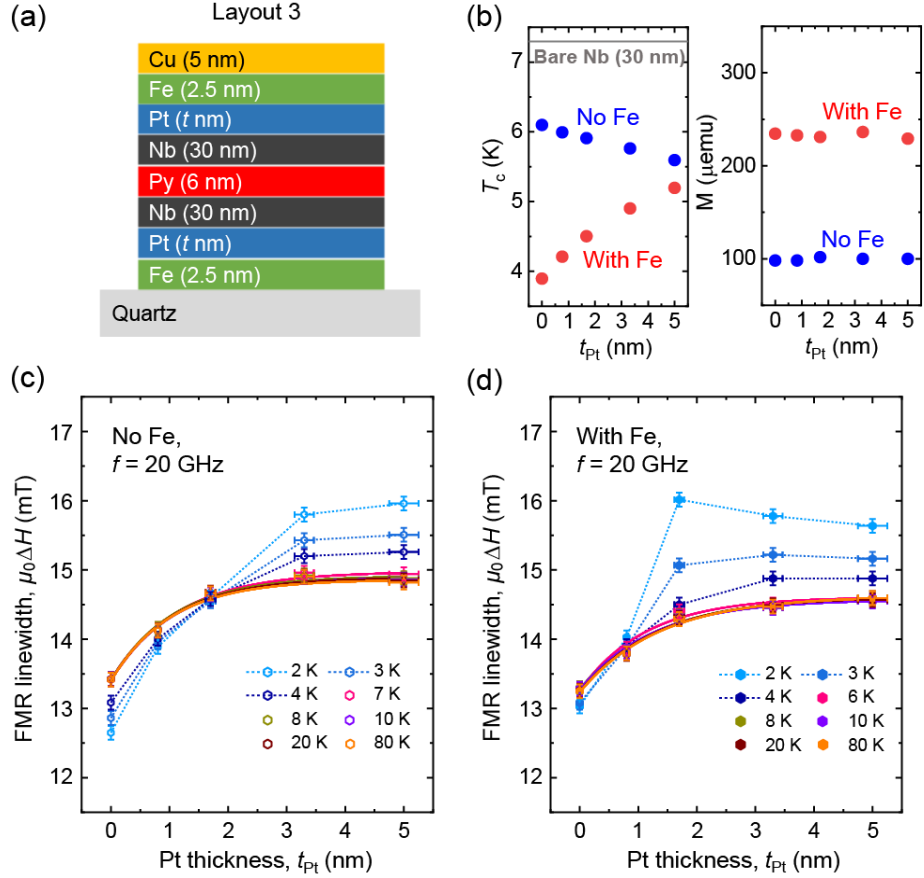

**FIG. 7.** (a) Layout of heterostructures with an additional Fe layer on Pt to study the role of a boosted internal exchange field in Pt on spin supercurrents generation.[90] (b) $T_c$ (left) and saturation magnetizations (right) as a function of Pt thickness for layout 2 [shown in Fig. 5(a)] and layout 3. (c) and (d) FMR linewidth as a function of Pt thickness for layout 2 and layout 3, respectively. The solid fitting curves are based on the spin pumping theory.

As shown in Fig. 7(a), a 2.5-nm-thick Fe layer is grown in contact with Pt with the intention to enhance the exchange field in Pt.[90] This is confirmed by the observation of an inverse proximity effect that causes an additional lowering of $T_c$ when the Fe layer is present [Fig. 7(b), left]. Because of the additional magnetic layer, the total magnetization of the structure increases [Fig. 7(b), right]; however, the independence on Pt thickness ($t_{Pt}$) is an indication that interface effects such as intermixing and interdiffusion are negligible. Fig. 7(c) and 7(d) compare the FMR linewidth *vs* $t_{Pt}$ when the additional Fe layer is absent (c) and present (d). Above $T_c$, the two graphs show similar trends, excluding that the growth of Fe impacts the spin sink quality, whereas below $T_c$ trends with $t_{Pt}$ are substantially different. When Fe is present, the linewidth does not simply saturate as $t_{Pt}$ increases above 4 nm, but reaches a maximum at $t_{Pt}$ = 1.5 nm. This suggests that the Fe effectively boosts $h_{ex}$ that can penetrate to the Nb/Pt interfaces, and the spin pumping efficiency is determined by both SOC and $h_{ex}$.

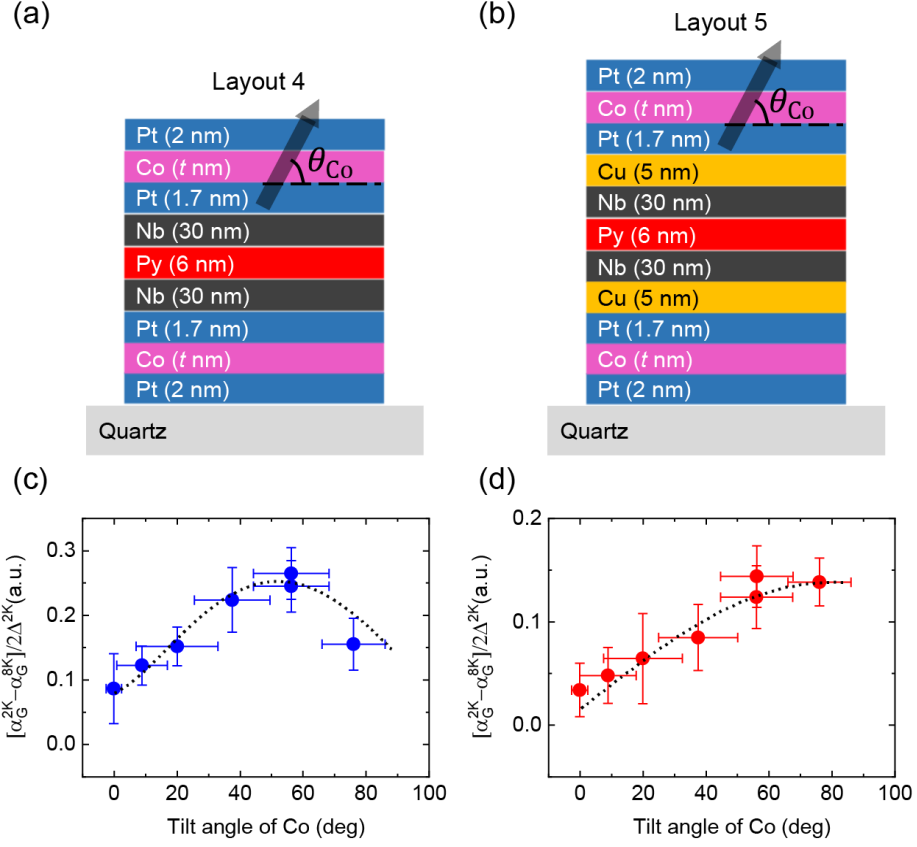

**FIG. 8.** (a) Layout 4 includes a Pt/Co/Pt stack to study the role of angle between the SOC field direction and the exchange field direction in spin supercurrent generation.[91] (b) Layout 5 includes an additional Cu spacer between the Nb and Pt layers to exclude any SOC-induced contribution to spin supercurrents generation. (c) and (d) Normalized damping difference across $T_c$ as a function of the tilt angle of Co magnetization. The dashed fitting curves are based on the spin-triplet proximity theory.

In the second set of experiments, the Pt spin sink is substituted with a Pt/Co/Pt stack [Fig. 8(a)], where Co serves as an internal source of an out-of-plane $h_{ex}$.[91] The angle of $h_{ex}$ is related to the magnetic anisotropy of Co and hence can be tuned by changing the Co layer thickness. Fig. 8(c) summarizes the results by plotting the superconductivity-induced change in Gilbert damping, renormalized by the superconducting gap, *vs* Co magnetization tilt angle. This normalization is necessary to compensate for the shrinking of the superconducting volume by Abrikosov vortex nucleation in the presence of an out-of-plane component of the stray field.[92] In these structures, another proximity-induced effect could contribute to triplets formation due to the magnetic non-collinearity of the two Co layers at the two sides of Nb. This effect, previously studied in Josephson junctions, depends on the relative angle between the two FMs sandwiching the superconductor and reaches a maximum value when they are orthogonal with each other. To subtract this contribution from any SOC-related contribution at the Nb-Pt interface, similar samples are measured in which additional Cu layers are inserted between the Nb and Pt layer [Fig. 8(b)]. The different angular dependence observed in the two sets of samples indicates competing effects that reach a maximum when $\theta_{Co} \sim 45°$. According to theory,[73] this angular dependence is consistent with a k-linear SOC term with Rashba symmetry.

## 4.4 Challenges and prospects of superconducting spintronics

Recently, Baek *et al.*[93] experimentally demonstrated STT switching that results in tenfold changes of the Josephson critical current in a nanopillar spin-valve Josephson junction. In this experiment the current-induced magnetization switching occurs in the normal state since the switching current is one order of magnitude larger than the Josephson critical current. Further reducing the switching current by using a magnetic material with lower magnetization, larger spin polarization, and lower damping could enable STT switching by spin-polarized quasiparticles. A recent experimental work reported highly spin polarized supercurrents with critical current densities exceeding $10^5$ A/cm$^2$ in $CrO_2$-based Josephson junctions.[94] Subsequent theory shows that this supercurrent can generate spin torques on an adjacent FM layer.[95] Typical current densities used to move domain walls via STT are of the order $10^7$ A/cm$^2$, still larger than the superconducting critical current density. SOT can lower this threshold value to levels that are compatible with superconductivity. It is also reported that the critical current density of type-II superconductor films can be enhanced by up to 30% via gate control.[96] A possible mechanism is surface nucleation and pinning of Abrikosov vortices. Voltage-controlled enhancements of the superconducting properties would also pave the way for the use of superconductors in magnetic switching experiments. A further point to consider in experimental design is the transmission of spin triplets through different materials. Our recent work shows a strong suppression of triplet supercurrents in the normal and superconducting states of Nb due to the enhanced spin scattering and lack of available equilibrium states, respectively.[97] Therefore, proper stack materials should be selected in spin triplet devices to ensure good spin-transfer efficiency.

In conventional spintronics, the current-induced torques are affected by epitaxial growth and crystalline symmetry[98-101]. The superconducting equivalent of the SOT has been studied in a number of theoretical works for different layout structures, such as Josephson junctions[102-105] and SC/FM bilayers.[106,107] We speculate that the symmetry of the system will also affect the symmetry of the torque[108] as it does for non-superconducting structures. To date, there is no experimental evidence of supercurrent-induced torques. In a recent work,[109] current-induced torques were characterised in superconductor-ferromagnet heterostructures, but no evidence of supercurrent-induced SOTs emerged from the study. This could be due to the absence of a triplet condensate due to the thickness of the superconducting layer exceeding the Cooper pairs correlation length.[84,110] Similar studies and thickness dependence characterizations are therefore highly desirable.

We note that the quantum coherence within a Josephson junction creates an addition potential energy that does not exist in the normal state, namely, the Josephson energy $U = (\Phi_0 I_c/2\pi)(1-\cos\Phi)$, which is associated with the phase difference $\Phi$ between the two S layers . In a spin-valve Josephson junction in which the critical current $I_c$ is determined by the relative alignment of magnetization $\cos(\theta)$—for example, in a S/F1/F2/S spin-valve Josephson junction—this energy is then a function of $\theta$. This implies that $\Phi$ can be used to directly control $\theta$ such that in a spin-valve structure, a FM free layer can be switched via the superconducting phase, provided that $U$ (~ 200 meV for 100 µA critical current) is large enough to overcome the reversal energies. Hence, the Josephson energy is an additional term that may assist superconducting-based STT processes, lowering switching energies and currents below $I_c$.

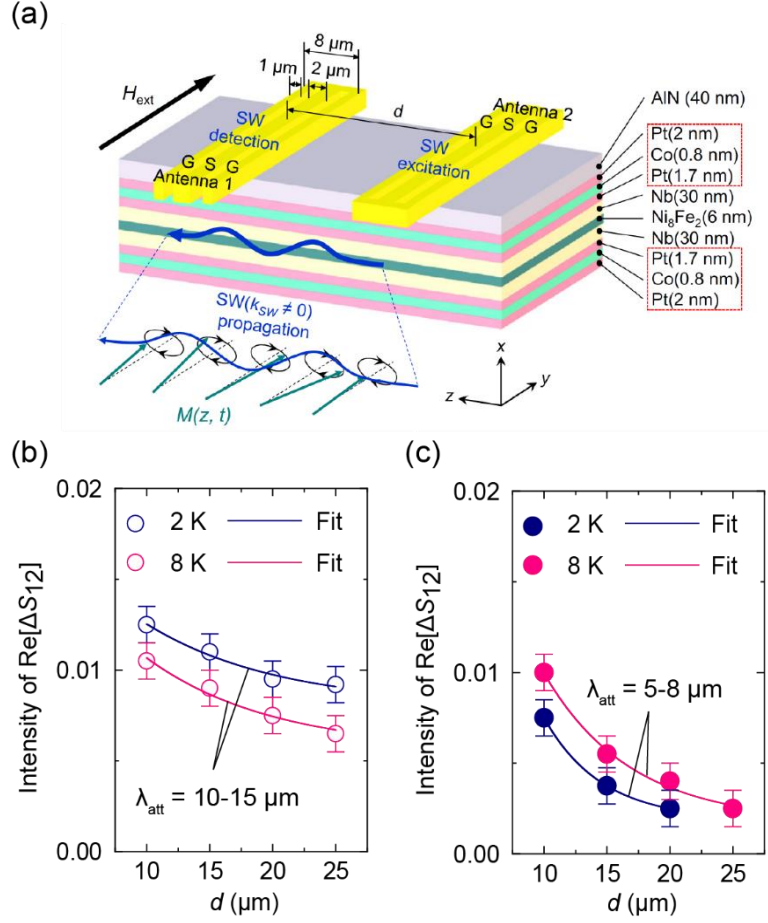

**FIG. 9.** (a) Schematic illustration of spin wave propagation on the proximity-engineered structure with the Pt-Co-Pt spin sink. When the middle NiFe layer is magnetized along the positive y direction, the spin wave propagation is along the positive z direction. (b) and (c) Real part of the spin wave transmission coefficient $\Delta S_{12}$ detected with a vector network analyser for samples without and with Pt-Co-Pt spin sink, respectively. The solid curves are fits to an exponential decay function to estimate the spin wave propagation length.[91]

The demonstration of efficient spin-charge conversion via the SHE is an important step toward energy-efficient magnetization switching. A recent study demonstrated that at SC/FM insulator interfaces, the conversion efficiency of magnons' spin into quasiparticles' charge via the inverse SHE is enhanced up to three orders of magnitude compared with that in the normal state, particularly when the interface superconducting gap matches the magnons spin accumulation.[59] The efficient spin-charge conversion at the superconducting transition point can be explained by two competing effects: superconducting coherence and exchange-field-modified quasiparticle relaxation, further pointing toward the important role of spin-split superconductivity in spin transmission processes. This result also suggests that antiferromagnetic insulators, with magnon excitation typically in the THz frequency range, which is comparable to the superconducting gap,[111] would allow efficient spin-charge conversion at temperatures much below $T_c$. In comparison, the inverse proximity effect from FM conductors hinders the formation of the coherence peak in the quasiparticle DOS,[112] which is manifested with an abrupt decrease of the inverse spin Hall voltage.[58]

The interaction between triplet Cooper pairs and magnons constitutes a fascinating research direction. Magnon-based devices have been explored as interconnecting elements in spin logic architectures, and achieving magnon modulation via the magnetic field[113] or electrical gating[114-116] is important for implementing magnonic logic functionalities. A recent study demonstrates that spin wave transmission within a Py layer between injector and detector antennas is modulated up to 40% by opening and closing a spin dissipation channel consisting of triplet correlations in a nearby Nb layer, as shown in Fig. 9(a).[91] Fig. 9(b) and (c) plot the real part of the transmission $\Delta S_{12}$ as a function of distance between the antennas for two sets of devices: one without (b) and one with (c) the exchange coupled Pt-Co-Pt spin sink in contact with Nb. While in the first type of devices, the transition to the superconducting state results in spin blocking, hence in a longer spin wave propagation length, in the second type the trend is inverted and the propagation length is suppressed by the formation of a spin triplet dissipation channel via proximity coupling to the spin sink.

## 5  Conclusions

This review is organized around strategies toward energy-efficient spintronics from the aspects of devices, materials, and physics. Starting from the scope of conventional spintronics, we briefly introduce the recent development of spintronic functionalities and devices, which would address outstanding issues such as static power dissipation and the von Neumann bottleneck in conventional CMOS technologies. Emergent materials with high charge-spin conversion efficiency are highlighted toward efficient magnetization switching via spin-orbit torque. In particular, we provide an overview into past and present activities related to superconducting spintronics, which offers new mediums to carry spin instead of charge, which may lead to massive reductions in Ohmic losses and support the development of novel spin-based devices. The exploration of superconducting spin currents is just emerging, demonstrating fascinating potential for novel spintronic device concepts. Going forward, fundamental parameters including spin-polarization, the conversion rate from singlet-to-triplet pairs, and triplet decay lengths and times, need to be determined experimentally and understood theoretically.

## Acknowledgements

We acknowledge funding from the EPSRC Programme Grant "Superspin" (no. EP/N017242/1) and the EPSRC International Network Grant "Oxide Superspin" (no. EP/P026311/1). Chiara Ciccarelli acknowledges support from the Royal Society.

## Data availability

Data sharing is not applicable to this article as no new data were created or analyzed in this study.